
\input phyzzx.tex


\def\np{Nucl. Phys.}
\def\pl{Phys. Lett.}

\def\cmp{Comm. Math. Phys.}
\def\ijmp{Int. J. Mod. Phys.}
\def\mpl{Mod. Phys. Lett.}

\def\phyrep{Phys. Rep.}

\def\sdg{Surveys in Diff. Geom.}

\def\fm{\phi_M}
\def\fmc{\phi_M^\dagger}
\def\fl{\phi_L}
\def\flc{\phi_L^\dagger}
\def\dfm{\partial \phi_M}
\def\dfmc{\partial \phi_M^\dagger}
\def\dfl{\partial \phi_L}
\def\dflc{\partial \phi_L^\dagger}
\def\psm{\psi_M}
\def\pmc{\psi_M^\dagger}
\def\psl{\psi_L}
\def\plc{\psi_L^\dagger}

\def\l{\lambda}

\tolerance=500000
\overfullrule=0pt
\Pubnum={US-FT-24/95}
\pubnum={US-FT-24/95}
\date={August, 1995}
\pubtype={}
\titlepage

\title{ Topological supergravity structure of non-critical
superstring theories}
\author{A. V. RAMALLO\foot{E-mail: ALFONSO@GAES.USC.ES},
S. ROY\foot{E-mail:ROY@GAES.USC.ES} and
J.M. SANCHEZ DE SANTOS \foot{E-mail: SANTOS@GAES.USC.ES}}
\address{Departamento de F\'\i sica de
Part\'\i culas, \break Universidad de Santiago, \break
E-15706 Santiago de Compostela, Spain.}

\abstract{ We obtain a bosonization prescription that allows to
represent the energy-momentum tensor and supersymmetry generators of
non-critical superstring theories with minimal matter as those of
topological supergravity. Superstrings with $N=1$ and $N=2$ world-sheet
supersymmetry are considered. The topological symmetry associated with
the topological supergravity representation is studied. It is shown, in
particular, that the compatibility of this topological structure with
the supersymmetry enhances the superconformal symmetry of the models
concerned.}

\endpage
\pagenumber=1
\sequentialequations
One of the main motivations to study topological field theories  was
the conjecture that (super)string theory may have a topological
phase
\REF\wittop{E. Witten \journal\cmp&117(88)353 \journal\cmp&118(88)411.}
\REF\bbrt{For a review see D. Birmingham, M. Blau, M.Rakowski
and G. Thompson \journal\phyrep&209(91)129.}[\wittop,\bbrt]. For this and
other reasons a lot of research effort has been directed to the study of
topological structures in non-critical string theories
\REF\witgrav{E. Witten\journal\np&B340(90)281 \journal\sdg&1(91)243.}
\REF\DW{R. Dijkgraaf and E. Witten\journal\np&B342(90)486.}
\REF\dij{R. Dijkgraaf, E. Verlinde and H. Verlinde
\journal\np&B352(91)59.;``Notes on topological string theory
and 2d quantum gravity", Proceedings of the Trieste spring
school 1990, edited by M.Green et al. (World Scientific,
Singapore,1991).}
\REF\gato{B. Gato-Rivera and A.M.
Semikhatov\journal\pl&B293(92)72 \journal\np&B408(93)133.}
\REF\BLNW{M. Bershadsky, W. Lerche, D. Nemechansky and N.
Warner\journal\np&B401(93)304.}
\REF\Mukhi{S. Mukhi and C. Vafa\journal\np&B407(93)667.}
\REF\PandaRoy{S. Panda and S. Roy\journal\pl&B317(93)533.}
[\witgrav\ - \PandaRoy]. The hope is
that the study of the topological aspects of string theory might provide
some clues that could help to understand the underlying symmetry
structure of (super)string theory.

Topological (super)gravity is believed to play an important role in the
hypothetical unbroken phase of (super)string theories. It is thus
important to identify the topological (super) gravity aspects of
(super)string theory. In ref.
\REF\LlatasRoy{P. M. Llatas and S. Roy\journal\pl&B342(95)66.}
[\LlatasRoy] the bosonic non-critical string theory
with minimal matter was mapped to the free field system introduced in
ref.
\REF\LPW{J.M.F. Labastida, M. Pernici and E.
Witten\journal\np&B310(88)611.}
[\LPW] to represent topological gravity. This mapping is achieved by
means of a convenient bosonization, generalizing the one in ref.
\REF\dist{J. Distler \journal\np&B342(90)523.}
[\dist], in
which the matter and Liouville fields are combined to form an odd copy
of the ghost sector, \ie\ a spin-two pair of commuting fields.
The resulting system is a topological ghost model, that has a
natural topological BRST symmetry. It is the
purpose of this paper to find a similar mapping for the superstring
case. Two different cases will be analyzed : the $N=1$
 Neveu-Schwarz-Ramond (NSR) and the $N=2$ non-critical superstrings.

Let us consider first the $N=1$ NSR non-critical superstring
\REF\PZ{A.M. Polyakov and A.B. Zamolodchikov\journal\mpl&A3(88)1213.}
\REF\DHK{J. Distler, Z. Hlousek and H. Kawai\journal\ijmp&A5(90)391.}
\REF\DiF{P. Di Francesco, J. Distler and D.
Kutasov\journal\mpl&A5(90)2135.}[\PZ, \DHK, \DiF]. The
components of the matter and Liouville supermultiplets  will be denoted
by $(\phi_M, \psi_M)$ and $(\phi_L, \psi_L)$ respectively, where
$\phi_M$ and $\phi_L$ are scalar fields and $\psi_M$ and $\psi_L$ are
Majorana fermions. The ghost sector of the theory contains the
anticommuting ghosts $b$ and $c$ with conformal weights $2$ and $-1$
together with the bosonic ghosts $\beta$ and $\gamma$ with conformal
weights $3/2$ and $-1/2$. We will concentrate ourselves in the analysis
of the holomorphic sector of the theory. The basic operator product
expansions (OPE's) of the fields will be taken as:
$$
\eqalign{
\phi_M(z)\,\phi_M(w)\,\sim&\,\phi_L(z)\,\phi_L(w)\,\sim\,
-{\rm log}(z-w)\cr
\psi_M(z)\,\psi_M(w)\,\sim&\,\psi_L(z)\,\psi_L(w)\,\sim
\,b(z)c(w)\,\sim\,\beta(z)\gamma(w)\,\sim\,
{1\over z-w}.\cr}
\eqn\uno
$$
The energy-momentum tensor $T$ of the theory can be represented as
$T\,=\,T^M\,+\,T^L\,+\,T^{\rm gh}$, where the contributions from
matter, Liouville and ghost fields are given by:
$$
\eqalign{
T^M\,&=\,-{1\over 2}\,(\partial\phi_M)^2\,+
iQ_M\partial^2\,\phi_M\,-
\,{1\over 2}\,\psi_M\partial\psi_M\cr
T^L\,&=\,-{1\over 2}\,(\partial\phi_L)^2\,+
iQ_L\partial^2\,\phi_L\,-
\,{1\over 2}\,\psi_L\partial\psi_L\cr
T^{\rm gh}\,&=\,-2b\partial c-\partial b c\,+\,
{3\over 2}\,\beta\partial\gamma\,+
\,{1\over 2}\,\partial\beta\gamma.\cr}
\eqn\dos
$$
In eq. \dos\ we have adopted a Coulomb gas representation of the matter
and Liouville sectors. $Q_M$ and $Q_L$ parametrize the matter and
Liouville background charges. Since the
total central charge must vanish and the central charge of the ghost
sector is $-15$, these quantities must satisfy the constraint
$Q_M^2\,+\,Q_L^2\,=-1$.  In eq. \dos, as well as in the following,
the products of fields should be understood as normal-ordered.
 The $N=1$ supersymmetry of this system is a
consequence of  the fact that there exists a dimension-$3/2$ fermionic
field (denoted by $T_F$) that verifies the basic OPE of the $N=1$
superconformal algebra:
$$
T_F(z)\,\,T_F(w)\,\sim\,{1\over 2}\,\,{T(w)\over z-w}.
\eqn\tres
$$
Given the representation \dos\ for $T$, it is easy to find the
expression for $T_F$. Indeed one can check that the operator
$$
T_F\,=\,{i\over 2}\,\partial\phi_M\psi_M\,+\,
{i\over 2}\,\partial\phi_L\psi_L\,+\,
Q_L\partial\psi_L\,+\,Q_M\partial\psi_M\,+\,
{1\over 2}\,\gamma b\,+\,
{3\over 2}\partial c \beta\,+\,c\partial\beta,
\eqn\cuatro
$$
satisfies eq. \tres. Under the action of $T_F$ the fields of the theory
are classified in supersymmetry doublets. One of such doublets
$(X,Y)$ is constituted by two operators $X$ and $Y$ of opposite
statistics and whose conformal weights differ by $1/2$
( $\Delta_Y=\Delta_X\,\, +\,\,{1\over 2}$ ). The action of $T_F$ on
$(X,Y)$ is given by:
$$
\eqalign{
T_F(z)\,\,X(w)\,\sim&\,\,{1\over 2}\,\, {Y(w)\over z-w}\cr\cr
T_F(z)\,\,Y(w)\,\sim&\,\,{\Delta_X X(w)\over (z-w)^2}+
{1\over 2}\,\, {\partial X(w)\over z-w}.\cr}
\eqn\cinco
$$
The ghost fields $b$, $c$, $\beta$ and $\gamma$ can be accommodated in
two of such $N=1$ doublets. Indeed one may check that eq. \cinco\ is
satisfied for $X=c$ and $Y=\gamma$ whereas the antighosts $\beta$ and
$b$ verify \cinco\ with $X=\beta$ and $Y=-b$. Moreover, the matter and
Liouville fields also form two supersymmetry doublets, which are
however  anomalous due to the existence of the background charges $Q_M$
and $Q_L$.

As it was pointed out above, the purpose of this paper is to find a
mapping from non-critical superstring theories to topological ghost
systems. Essentially we want to show that one can combine the matter and
Liouville fields in such a way that they can be regarded as a
topological copy of the superstring ghost sector. By a topological copy
we mean a series of fields having the same conformal dimensions as the
ghosts $(b,c)$ and $(\beta,\gamma)$ but obeying opposite statistics.
Accordingly we want to represent the matter + Liouville sector of the
superstring in terms of two pairs of fields $(B,C)$ and $({\cal B},
\Gamma)$. The fields $B$ and $C$ (${\cal B}$ and $\Gamma$) are bosonic
(fermionic) and they have the same conformal dimensions as their
lower case counterparts (\ie\  $\Delta_B\,=\,2$, $\Delta_C\,=\,-1$,
$\Delta_{\cal B}\,=\,3/2$ and $\Delta_\Gamma\,=\,-1/2$). The resulting
topological ghost system is nothing but the free field representation of
topological supergravity
\REF\Hughes{D. Montano, K. Aoki and J.
Sonnenschein\journal\pl&B247(90)64;
J. Hughes and K. Li\journal\pl&B261(91)269.}
[\Hughes], which is the supersymmetric
generalization of the ghost system used in ref. [\LPW] to represent
topological gravity.

Supersymmetry will play a central role in our approach and thus it is
natural to require that its generator $T_F$ should be realized locally in
terms of the new fields $B$, $C$, ${\cal B}$ and $\Gamma$. The best way
to achieve this locality requirement is by demanding that
$B$, $C$, ${\cal B}$ and $\Gamma$ belong to supersymmetry doublets.
Taking into account the conformal weights and statistics of these
fields, the only possible doublets that one can form with them are
$(C,\Gamma)$ and $({\cal B},B)$. Let us adopt a vector notation for the
fields of the matter and Liouville sectors. We shall assemble the scalar
and Majorana fields in a two component vector, \ie\ we define
$\vec \phi\,=\,(\phi_M,\phi_L)$ and $\vec \psi\,=\,(\psi_M,\psi_L)$.
Similarly the background charges $Q_M$ and $Q_L$ will be considered as
the two components of the vector $\vec Q$
(\ie\  $\vec Q\,=\,(Q_M,Q_L)$).

In order to represent the $(C,\Gamma)$ and $({\cal B},B)$ doublets in
terms of the fields $\vec \phi$ and $\vec \psi$ we follow the manifest
supersymmetric bosonization of refs.
\REF\Martinec{E. Martinec and G. Sotkov\journal\pl&B208(88)240.}
\REF\Takama{M. Takama\journal\pl&B210(88)153.}
[\Martinec, \Takama]. Let us first
apply this bosonization formalism to the $(C,\Gamma)$  doublet. The
field $C$ in this approach is represented as an exponential of
$\vec \phi$, \ie\ as an expression of the form
$C\,=\,e^{\vec \mu\cdot\vec\phi}$ with $\vec\mu$ a complex constant
numerical vector. Once the form of $C$ is known, its
supersymmetric partner $\Gamma$ is obtained by acting with $T_F$ and
comparing the resulting expression with the first equation in \cinco.
After a simple calculation one gets:
$$
C\,=\,e^{\vec \mu\cdot\vec\phi}
\,\,\,\,\,\,\,\,\,\,\,\,\,\,\,\,\,\,\,\,\,\,\,\,\,\,
\Gamma\,=\,-i\vec\mu\cdot\vec\psi\,e^{\vec \mu\cdot\vec\phi}.
\eqn\seis
$$
The $({\cal B},B)$ form a doublet which is conjugate to $(C,\Gamma)$.
This means that the only non-vanishing OPE's among these fields are:
$$
{\cal B}(z)\,\Gamma(w)\,\sim\,B(z)\,C(w)\,\sim\,{1\over z-w}.
\eqn\siete
$$
By inspecting the form of $\Gamma$ one realizes that  there is a
chance  to reproduce the correct OPE ${\cal B}(z)\,\Gamma(w)$ only when
${\cal B}$ and $\Gamma$ are given by a similar expression.
 Accordingly we consider the following ansatz for  ${\cal B}$ and $B$:
$$
{\cal B}\,=\,\vec\rho\cdot\vec\psi\,e^{-\vec \mu\cdot\vec\phi}
\,\,\,\,\,\,\,\,\,\,\,\,\,\,\,\,\,\,\,\,\,\,\,\,\,\,
B\,=\,i[\,\vec\mu\cdot\vec\psi\,\vec\rho\cdot\vec\psi\,+\,
\vec\rho\cdot\partial\vec\phi\,]e^{-\vec \mu\cdot\vec\phi},
\eqn\ocho
$$
where $\vec\rho$ is a new numerical vector. Notice that the expression
of $B$ can be  obtained by acting  with $T_F$ on ${\cal B}$. Actually
the OPE's of eqs. \cinco\ and \siete\ impose conditions on the inner
products of the vectors $\vec \mu$, $\vec\rho$ and $\vec Q$. These
conditions are:
$$
\eqalign{
&\vec \mu^{\,\,2}\,=\,\vec \rho^{\,\,2}\,=\,0
\,\,\,\,\,\,\,\,\,\,\,\,\,\,\,\,\,\,\,\,\,\,\,\,\,\,
\vec\rho\cdot\vec\mu\,=\,i\cr
&\,\,\,\,\,\,\,\,\,\,
\vec Q\cdot\vec\mu\,=\,i
\,\,\,\,\,\,\,\,\,\,\,\,\,\,\,\,\,\,\,\,\,\,\,\,\,\,
\vec Q\cdot\vec\rho\,=\,-{1\over 2}.\cr}
\eqn\nueve
$$
Eq. \nueve\ can be easily solved for $\vec\mu$ and $\vec\rho$. In order
to find this solution, let us parametrize the background charges $Q_M$
and $Q_L$ as:
$$
Q_M\,=\,{1\over 2}\,({1\over \lambda}\,-\,\lambda\,)
\,\,\,\,\,\,\,\,\,\,\,\,\,\,\,\,\,\,\,\,
Q_L\,=\,{i\over 2}\,(\lambda\,+\,{1\over \lambda}\,),
\eqn\diez
$$
where $\lambda$ is a constant. Notice that the condition
$Q_M^2+Q_L^2=-1$ is automatically satisfied. The parameter $\lambda$ is
related to the central charge of the matter sector. For example, for
$(p,q)$ minimal superconformal matter $\lambda\,=\,\sqrt{q\over p}$. In
terms of $\lambda$ the solution of eq. \nueve\ is:
$$
\vec\mu\,=(i\lambda,\lambda)
\,\,\,\,\,\,\,\,\,\,\,\,\,\,\,\,\,\,\,\,\,\,\,\,\,\,
\vec\rho\,=\,({1\over 2\lambda},{i\over 2\lambda}).
\eqn\once
$$
Using these values of $\vec\mu\,$ and $\vec\rho\,$ in our bosonization
formulas (eqs. \seis\ and \ocho), we can get the explicit expressions of
the fields $B$, $C$, ${\cal B}$ and $\Gamma$. The result is:
$$
\eqalign{
B\,=&\,i[-\psi_M\psi_L\,+\,{1\over 2\lambda}\,
(\partial\phi_M\,+\,i\partial\phi_L\,)]\,
e^{-i\lambda(\phi_M-i\phi_L)}\cr\cr
C\,=&e^{i\lambda(\phi_M-i\phi_L)}\cr\cr
{\cal B}\,=&\,{1\over 2\lambda}\,
(\psi_M+i\psi_L)\,e^{-i\lambda(\phi_M-i\phi_L)}\cr\cr
\Gamma\,=&\lambda(\psi_M-i\psi_L)\,
e^{i\lambda(\phi_M-i\phi_L)}.\cr}
\eqn\doce
$$
It is now easy to express $T$ and $T_F$ in terms of the new fields.
First of all it is straightforward to check that the matter and
Liouville contributions to the energy-momentum tensor are given by:
$$
T^M\,+T^L\,=\,2B\partial C \,+\,\partial B C\,-
\,{3\over 2}\,{\cal B}\partial \Gamma\,-\,
{1\over 2}\,\partial {\cal B}\Gamma,
\eqn\trece
$$
Moreover, the supersymmetry generator $T_F$ can be written as:
$$
T_F\,=\,{1\over 2}\, \Gamma B\,+\,
{3\over 2}\,\partial C{\cal B}\,+\,C\partial{\cal B}\,+\,
{1\over 2}\,\gamma b\,+\,
{3\over 2}\partial c \beta\,+\,c\partial\beta.
\eqn\catorce
$$
A simple comparison of the ghost and matter+Liouville contributions to
$T$ and $T_F$ reveals the existence of a clear symmetry relating them.
This symmetry is in fact a topological BRST symmetry as we shall check
below. The current associated with the generator of this symmetry can
be taken as:
$$
Q\,=\,bC\,-\,\beta\Gamma .
\eqn\quince
$$
It is a simple exercise to verify that $T$ and $T_F$ are BRST invariant
\ie\ that they are invariant under the action of the zero-mode of $Q$.
Actually it is interesting to point out that the relative coefficient
of the two terms of $Q$ (including the sign) is uniquely fixed by the
BRST invariance of $T_F$, which is a condition that we must require in
order to have a topological symmetry compatible with the supersymmetry
of the model. The BRST current $Q$ endows the NSR superstring with the
structure of a topological Conformal Field Theory. In such theories the
energy-momentum tensor $T$ is $Q$-exact and its  BRST ancestor is
usually denoted by $G$. The local relation between $G$ and $T$ is
determined by the OPE:
$$
Q(z)\,G(w)\,\sim\,{d\over (z-w)^3}+ {R(w)\over (z-w)^2}
+{T(w)\over z-w},
\eqn\dseis
$$
where $d$ is a c-number anomaly and $R$ is a $U(1)$ current.
In our case, \ie\ when $Q$ is given by eq. \quince, $\,d=-1$ while $G$
and $R$ are:
$$
\eqalign{
G\,=&\,c\partial B\,+\,2\partial cB\,-\,{1\over 2}\,
\gamma\partial {\cal B}\,-
\,{3\over 2}\, \partial \gamma{\cal B}\cr
R\,=&\,cb\,+\,2BC\,+\,{1\over 2}\, \beta\gamma
\,+\,{3\over 2}\, \Gamma{\cal B},\cr}
\eqn\dsiete
$$
It is easy to check that $T$, $G$, $Q$ and $R$ close a topologically
twisted $N=2$ superconformal algebra
\REF\LVW{W. Lerche, C. Vafa and N.P. Warner
\journal\np&B324(89)427.}
\REF\EY{T. Eguchi and S.-K. Yang \journal\mpl&A4(90)1653;
T. Eguchi, S. Hosono and S.-K. Yang \journal\cmp&140(91)159.}
 [\LVW, \EY]. Let us now try to find the
operator algebra closed by $T_F$ and $T$, $G$, $Q$ and $R$. This algebra
determines how the topological symmetry and the supersymmetry are
interrelated. It is interesting to remember here that the $U(1)$
current $R$ defines a grading which is characteristic of the
topological symmetry we are dealing with. In fact all the generators of
the topological algebra should have a well-defined $R$-charge. From the
explicit expression of $R$ in eq. \dsiete\ it follows that the
$R$-charges of the fields $b$, $c$, $\beta$, $\gamma$, $B$, $C$,
${\cal B}$ and  $\Gamma$ are respectively $-1$, $+1$, $-1/2$, $+1/2$,
$-2$, $+2$, $-3/2$ and $+3/2$. By a simple counting
 one easily concludes that  there are  two types of terms in $T_F$
with $R$-charges $\pm 1/2$. Actually the topological $U(1)$ current $R$
induces a splitting of $T_F$ in two pieces. Let us write
$T_F\,=\,{1\over 2}\, (T_F^{+}\,+\,T_F^{-}\,)$, where $T_F^{+}$ and
$T_F^{-}$ have
$R$-charges $+1/2$ and $-1/2$ respectively and whose explicit expressions
are given by:
$$
T_F^+\,=\,3\,\partial C{\cal B}\,
+\,2C\partial{\cal B}\,+\,
3\partial c \beta\,+\,2c\partial\beta
\,\,\,\,\,\,\,\,\,\,\,\,\,\,\,\,\,\,\,\,\,\,\,\,\,\,\,\,
T_F^-\,=\, \Gamma B\,+\,\gamma b
\eqn\docho
$$
It is interesting to notice that the splitting of eq. \docho\ is quite
special. In fact $T_F^{\pm}$ are the generators of an $N=2$
superconformal algebra. This fact may be verified by computing the
operator algebra closed by $T_F^{\pm}$. One of the relations in this
algebra is:
$$
T_F^+(z)\,T_F^-(w)\,\sim\,{J(w)\over (z-w)^2}\,+\,
{T(w)\,+{1\over 2}\,\partial  J(w)\over z-w},
\eqn\dnueve
$$
where $J$ is the $U(1)$ current associated to the $N=2$ superconformal
algebra which is given by:
$$
J\,=\,-2cb-2BC-3\beta\gamma-3\Gamma{\cal B}.
\eqn\veinte
$$
We thus see that the original $N=1$ supersymmetry of the string is
promoted to an $N=2$ superconformal symmetry. This enhancement of the
supersymmetry is essential in order to make topological symmetry and
supersymmetry compatible. In fact, the hidden $N=2$ superconformal
symmetry of the NSR ghost sector  was found some time
ago in ref.
\REF\FMS{D. Friedan, E. Martinec and S. Shenker
\journal\np&B271(86)93.}
[\FMS]. The topological symmetry exhibited by our
bosonization extends this $N=2$ supersymmetry to the matter+Liouville
sector.

Under the $N=2$ supersymmetry just uncovered, it should be possible to
arrange all operators of the theory in $N=2$ supermultiplets. In general
 such a supermultiplet is composed by four fields
$(X,Y^+,Y^-,Z)$ whose conformal weights are related as
$\Delta_{Y^{\pm}}\,=\,\Delta_X\,+\,1/2$ and
$\Delta_Z\,=\,\Delta_X\,+\,1$. The action of $T_F^{\pm}$ on the fields
of the supermultiplet is given by:
$$
\eqalign{
T_F^{\pm}(z)\,X(w)\,\sim&\,\mp{Y^{\pm}(w)\over z-w}\cr
T_F^{+}(z)\,Y^+(w)\,\sim&\,\,T_F^{-}(z)\,Y^-(w)\,\sim\,0\cr
T_F^{\pm}(z)\,Y^{\mp}(w)\,\sim&\,\pm\,
{\Delta_X X(w)\over (z-w)^2}\,+\,
{Z(w)\pm{1\over 2}\partial X(w)\over z-w}\cr
T_F^{\pm}(z)\,Z(w)\,\sim&\,\,{\Delta_{Y^{\pm}}Y^{\pm}(w)
\over (z-w)^2}\,+{1\over 2}\,{\partial Y^{\pm}(w)\over z-w}.
\cr}
\eqn\vuno
$$
An example of such an $N=2$ multiplet is the one constituted by the
generators of the $N=2$ superconformal algebra
$(J,T_F^{+},T_F^{-},T)$. The BRST ancestor of $T$, which we called
$G$ in eq. \dsiete, also belongs to  an $N=2$ multiplet. Let us denote
the members of this multiplet as $(g, G_B^+,G_B^-, G)$. The operators
$g$ and $G$ are fermionic whereas  $ G_B^+$ and $ G_B^-$ are bosonic.
Their conformal weights are  $\Delta_g\,=\,1$,
$\Delta_{G_B^{\pm}}\,=\,3/2$ and $\Delta_G\,=\,2$. Using eqs. \dsiete\
and \docho\ one can easily obtain the expressions of $g$ and
$G_B^{\pm}$. The result is:
$$
g\,=\,3{\cal B}\gamma-2c B
\,\,\,\,\,\,\,\,\,\,\,\,\,\,\,\,\,\,\,
G_B^+\,=\,-3\partial c{\cal B}\,-\,2c\partial {\cal B}
\,\,\,\,\,\,\,\,\,\,\,\,\,\,\,\,\,\,\,
G_B^-\,=\,B\gamma.
\eqn\vdos
$$
Acting the with BRST current $Q$ on $(g, G_B^+,G_B^-, G)$ one generates
$(J,T_F^{+},T_F^{-},T)$ as the residue of the single pole singularity.
The action of $Q$ on $G$ is displayed in eq. \dseis, whereas the action
of $Q$ on $g$ and $G_B^{\pm}$ is given by:
$$
\eqalign{
Q(z)\,g(w)\,\sim&\,-{1\over (z-w)^2}\,+\,{J(w)\over z-w}\cr
Q(z)\,G_B^+(w)\,\sim&-{R_F(w)\over (z-w)^2}\,-\,
{T_F^{+}(w)\over z-w}\cr
Q(z)\,G_B^-(w)\,\sim&-\,{T_F^{-}(w)\over z-w}.\cr}
\eqn\vtres
$$
Notice that, in particular, eq. \vtres\ shows that $J$ and $T_F^{\pm}$
are BRST-exact operators. In eq. \vtres\ $R_F$ is a dimension-$1/2$
fermionic current whose explicit expression is:
$$
R_F\,=\,2\beta c+3{\cal B}C.
\eqn\vcuatro
$$
Supermultiplets with four fields are not the only possibility to
arrange the operators of a theory with $N=2$ supersymmetry. We can also
have chiral multiplets constituted by two fields $(X,Y^-)$ that under
the action of $T_F^{\pm}$ behave as follows:
$$
\eqalign{
T_F^{-}(z)\,X(w)\,\sim &\,{Y^-(w)\over z-w}\cr
T_F^{+}(z)\,X(w)\,\sim &\,\,T_F^{-}(z)\,Y^-(w)\,\sim\,0\cr
T_F^{+}(z)\,Y^{-}(w)\,\sim &\,\,
{2\Delta_X X(w)\over (z-w)^2}\,+\,
{\partial X(w)\over z-w}.\cr}
\eqn\vcinco
$$
The ghost fields $(c,\gamma)$ and the corresponding antighosts
$(\beta, -b)$, together with their topological partners
$(C,\Gamma)$ and  $({\cal B}, B)$, are examples of chiral multiplets.
It turns out that the BRST current $Q$ also  belongs to a chiral
multiplet. It may be checked that there exists a bosonic
dimension-$1/2$ operator $Q_B$ such that the fields $(Q_B,Q)$  satisfy
eq. \vcinco. This supersymmetric partner of the BRST current is:
$$
Q_B\,=\,-C\beta.
\eqn\vseis
$$
There is still another chiral multiplet formed by the fermionic
operator $R_F$ and the particular combination $R-{1\over 2}J$ of the
$U(1)$ currents $R$ and $J$. However the action of $T_F^{\pm}$ on
$(R_F,R-{1\over 2}J)$ is anomalous (recall that $R$ is anomalous under
the action of $T$). In fact one can check that instead of the first
equation of \vcinco\ with $X=R_F$ and $Y^-=R-{1\over 2}J$ one has:
$$
T_F^-(z)\,R_F(w)\,\sim\,{1\over (z-w)^2}\,+\,
{R(w)-{1\over 2}J(w)\over z-w}.
\eqn\vsiete
$$
These two chiral multiplets are connected by means of the BRST
current $Q$. Indeed one can verify that the operators $(Q_B,Q)$ can be
obtained by acting  with $Q$ on the fields $(R_F,R-{1\over 2}J)$ . One
has, for example, that:
$$
Q(z)R_F(w)\,\sim\,{Q_B(w)\over z-w}.
\eqn\vocho
$$

The twelve operators introduced so far close an algebra which has  a
topologically twisted $N=2$ algebra (obeyed by $T$, $G$, $Q$ and $R$)
and an untwisted $N=2$ superconformal algebra (generated by $T$,
$T_F^+$, $T_F^-$ and $J$) as
subalgebras. The compatibility between the topological and
superconformal symmetries is reflected in the fact that all the
generators of the algebra have well-defined charges with respect to the
two $U(1)$ currents $R$ and $J$ of these two types of symmetries. In
fact one can verify that the operators $J$, $T_F^+$,  $T_F^-$, $T$, $g$,
$G_B^+$, $G_B^-$, $G$, $Q_B$, $Q$, $R_F$ and $R$ have $R$-charges
 $0$, $+1/2$, $-1/2$, $0$, $-1$, $-1/2$, $-3/2$, $-1$,
$3/2$, $1$, $+1/2$ and $0$ respectively, whereas the $J$-charges of these
same fields are $0$, $+1$, $-1$, $0$, $0$,  $+1$, $-1$, $0$, $+1$, $0$,
$+1$ and $0$. The OPE's of $R$ and $J$ with the generators of the
algebra always contain a single pole singularity dictated by these
charges. In some of these OPE's, however, higher order
poles could appear. Two examples of this anomalous
 behaviour under the action of the
$R$ and $J$ currents are:
$$
\eqalign{
R(z)\,T_F^+(w)\,\sim&\,{R_F(w)\over (z-w)^2}\,+\,{1\over 2}\,
{T_F^+(w)\over z-w}\cr
J(z)\,G(w)\,\sim&\,{g(w)\over (z-w)^2}.\cr}
\eqn\vnueve
$$
Other interesting OPE's of the algebra are:
$$
\eqalign{
G(z)\,R_F(w)\,\sim&\,-{G_B^+(w)\over z-w}\cr
G(z)\,Q_B(w)\,\sim&\,\,{1\over 2}\, {R_F(w)\over (z-w)^2}\,+\,
{\partial R_F(w)-T_F^+(w)\over z-w}\cr
G_B^+(z)\,R_F(w)\,\sim&\,G_B^+(z)\,Q_B(w)\,\sim\,0\cr
G_B^-(z)\,R_F(w)\,\sim&\,\,{1\over 2}{g(w)\over z-w}\cr
G_B^-(z)\,Q_B(w)\,\sim&\,\,{1\over (z-w)^2}\,+\,
{R(w)+{1\over 2} J(w)\over z-w}.\cr}
\eqn\treinta
$$

It would be interesting to relate the topological algebra just
described to some superconformal symmetry. Usually, in order to relate
topological and superconformal algebras one has to perform a twisting
in their generators. It turns out that our topological algebra can be
related to a small $N=4$ superconformal symmetry
\REF\Ademollo{M. Ademollo et
al.\journal\pl&B62(76)105\journal\np&B111(76)77.}
[\Ademollo]. This $N=4$
supersymmetry can be regarded as two $N=2$ supersymmetries that share
the same $U(1)$ current. The generators of the two $N=2$
supersymmetries will be denoted by $G_i^{\pm}$ for $i=1,2$ whereas the
common $U(1)$ current will be called $J_0$. The mixing of the two $N=2$
supersymmetries requires the introduction of two extra currents (denoted
by $J^{++}$ and $J^{--}\,$) which, together with $J_0$, close an affine
$SU(2)$ algebra ($J_0$ is the Cartan generator of this algebra).
Denoting the twisted energy-momentum tensor by $\tilde T$, the $N=4$
generators are:
$$
\eqalign{
G_1^+\,=&\,T_F^+\,-\,\partial R_F
\,\,\,\,\,\,\,\,\,\,\,\,\,\,\,
G_1^-\,=\,T_F^-
\,\,\,\,\,\,\,\,\,\,\,\,\,\,\,
G_2^+\,=\,Q
\,\,\,\,\,\,\,\,\,\,\,\,\,\,\,
G_2^-\,=\,G\,+\,{1\over 2}\,\partial g\cr
J_0\,=&\,R\,+\,{1\over 2}\,J
\,\,\,\,\,\,\,\,\,\,\,\,\,\,\,
J^{++}\,=\,Q_B
\,\,\,\,\,\,\,\,\,\,\,\,\,\,\,
J^{--}\,=-\,G_B^{-}\cr
\tilde T\,=&\,T\,-\,{1\over 2}\partial (R\,-\,{1\over 2}\,J).\cr}
\eqn\tuno
$$
Notice that the $N=4$ algebra has fewer generators than our topological
algebra (eight versus twelve). The central charge of $\tilde T$ is
$-6$. This is the Virasoro anomaly of an $N=2$ ghost multiplet. In fact
the conformal weights with respect to $\tilde T$ of the fields $(b,c)$,
$(\beta, \gamma)$, $(B,C)$ and $({\cal B}, \Gamma)$ are $(1,0)$,
$(1/2,1/2)$, $(1/2,1/2)$ and $(0,1)$ respectively. Given these
conformal dimensions one can form two conjugate $N=2$ multiplets:
$(c, -C, \gamma, -\Gamma)$ and $({\cal B}, \beta, B,-b)$. It can be
shown that, arranged in this way, these fields behave as in eq. \vuno\
with respect to the superconformal generators
$G_2^+ + {1\over 2}\,G_1^+$ and $G_1^- + {1\over 2}\,G_2^-$.  This means
that we have embedded the $N=1$ NSR string in an $N=2$ ghost system.

Let us now consider the $N=2$ non-critical superstring [\DHK]. In this
case the matter and Liouville sectors can be regarded as the
complexification of the $N=1$ NSR superstring. Therefore, along with
the bosonic and fermionic fields
$\vec \phi\,=\,(\phi_M,\phi_L)$ and $\vec \psi\,=\,(\psi_M,\psi_L)$, we
must deal with their hermitian conjugates
$\vec \phi^{\,\,\dagger}\,=\,(\phi_M^{\dagger},\phi_L^{\dagger})$ and
$\vec \psi^{\,\,\dagger}\,=\,(\psi_M^{\dagger},\psi_L^{\dagger})$. The
ghost sector now contains a pair of anticommuting fields
$(\tilde b, \tilde c)$ with conformal weights $(1,0)$, two commuting
pairs $(\beta_+, \gamma_-)$ and $(\beta_-, \gamma_+)$  with
conformal weights $(3/2, -1/2)$ and a pair $(b,c)$ of anticommuting
ghosts with conformal weights $(2,-1)$. The basic OPE's among these
fields are:
$$
\eqalign{
&\phi_M(z)\,\phi_M^{\dagger}(w)\,\sim\,
\phi_L(z)\,\phi_L^{\dagger}(w)\,\sim\,
-{\rm log}(z-w)\cr
&\psi_M(z)\,\psi_M^{\dagger}(w)\,\sim\,
\psi_L(z)\,\psi_L^{\dagger}(w)\,\sim\,
{1\over z-w}\cr
&\tilde b(z)\tilde c(w)\,\sim\,\beta_+(z)\gamma_-(w)\,\sim\,
\beta_-(z)\gamma_+(w)\,\sim\,b(z)c(w)\,\sim\,{1\over z-w}.\cr}
\eqn\xuno
$$
The generators of the $N=2$ world-sheet superconformal symmetry in the
matter+Liouville sector are given by:
$$
\eqalign{
T^M\,+\,T^L\,=&\, -\partial \vec \phi\cdot \partial\vec \phi^{\,\,\dagger}\,+\,
i\vec Q\cdot (\partial^2\vec \phi\,+\,\partial^2\vec
\phi^{\,\,\dagger}\,)\,-\, {1\over 2}\,
\vec\psi\cdot\partial\vec\psi^{\,\,\dagger}\, -\, {1\over 2}\,
\vec\psi^{\,\,\dagger}\cdot\partial\vec\psi\cr
 T_F^{ M,+}\,+\,T_F^{ L,+}\,=&\, i\partial
\vec\phi\cdot\vec\psi^{\,\,\dagger}\, +\,2\vec
Q\cdot\partial\vec\psi^{\,\,\dagger}\cr
T_F^{ M,-}\,+\,T_F^{ L,-}\,
=&\, i\partial \vec\phi^{\,\,\dagger}\cdot\vec\psi\,
+\,2\vec Q\cdot\partial\vec\psi\cr
J^M\,+\,J^L\,=&\,\vec\psi^{\,\,\dagger}\cdot\vec\psi\,
+\,2i\vec Q\cdot\partial\vec\phi\,
-\,2i\vec Q\cdot\partial\vec\phi^{\,\,\dagger},\cr}
\eqn\xdos
$$
whereas in the ghost sector the $N=2$ supersymmetry is generated by:
$$
\eqalign{
T^{\rm gh}\,=&-\,2b\partial c-\partial b c\,-\,\tilde b\partial\tilde c\,+\,
{3\over 2}\,\beta_+\partial\gamma_-\,+\,{1\over 2}\partial\beta_+\gamma_-\,+\,
{3\over 2}\,\beta_-\partial\gamma_+\,+\,{1\over 2}\partial\beta_-\gamma_+\cr
T_F^{{\rm gh},\pm}\,=&\mp b\gamma_{\pm}\,+\,\partial\gamma_{\pm}\tilde b\,+\,
{1\over 2}\,\gamma_{\pm}\partial\tilde b\,\pm\,{3\over 2}\,
\beta_{\pm}\partial c\,\pm\partial\beta_{\pm}c\,+\,\tilde c\beta_{\pm}\cr
J^{\rm gh}\,=&\,\gamma_+\beta_-\,-\, \beta_+\gamma_-
\,-\,\partial(\tilde b c).\cr}
\eqn\xtres
$$
The $N=2$ supersymmetry classifies the ghost fields in multiplets.
Indeed, one can check that eq. \vuno\ is satisfied for
$(X, Y^+, Y^-, Z)\,=\,(c, \gamma_+, \gamma_-, \tilde c)$. Similarly the
$N=2$ antighost multiplet is
$(X, Y^+, Y^-, Z)\,=\,(\tilde b, -\beta_+, \beta_-, b)$.
Notice that now the ghost central charge is equal to $-6$. This means
that the vector $\vec Q\,=\,(Q_M, Q_L)$ in eq. \xdos\ must satisfy
$\vec Q^2\,=\,Q_M^2\,+Q_L^2\,=0$. We shall not consider here the case
 $\vec Q\,=0$, which corresponds to the critical $N=2$ string. In fact,
as we did for the $N=1$ case in eq. \diez, we shall now parametrize $Q_M$
and $Q_L$ as follows:
$$
Q_M\,=\,{1\over 4\lambda}
\,\,\,\,\,\,\,\,\,\,\,\,\,\,\,\,\,
Q_L\,=\,{i\over 4\lambda}.
\eqn\xcuatro
$$
In analogy with what we have done in the $N=1$ NSR string, we want to
show how the matter+Liouville sector of the $N=2$ superstring can be
represented as a topological copy of the ghost sector. Accordingly, let
us consider two pairs $(\tilde B,\tilde C)$ and $(B,C)$ of commuting
fields with conformal weights $(1,0)$ and $(2,-1)$ respectively
together with another two pairs of anticommuting fields
$({\cal B}^+, \Gamma^-)$ and $({\cal B}^-, \Gamma^+)$ each of which
having conformal dimensions $(3/2, -1/2)$. The OPE's among these fields
will be taken to be:
$$
\tilde B(z)\tilde C(w)\,\sim\,
{\cal B}^+(z)\Gamma^-(w)\,\sim\,
{\cal B}^-(z)\Gamma^+(w)\,\sim\,
 B(z) C(w)\,\sim\,{1\over z-w}.
\eqn\xcinco
$$
In order to extract these new fields from the matter+Liouville sector,
the realization of the $N=2$ symmetry of the model in terms of them will
be our guiding principle. In complete parallel with the ghost sector, we
shall distribute the new fields in two $N=2$ multiplets. The content of
one of these multiplets will be
$(X, Y^+, Y^-, Z)\,=\,(C, \Gamma^+,\Gamma^-, \tilde C)$ while its
conjugate multiplet will be formed by  $(\tilde B, {\cal B}^+, -{\cal
B}^-, B)$. Let us adopt the following ansatz for the fields $C$ and
$\tilde B$ (\ie\ for the components of the multiplets with the lowest
dimension):
$$
C\,=\,e^{\vec \mu\cdot\vec\phi\,+\,
\vec\mu^{\,\,\dagger}\cdot\vec\phi^{\,\,\dagger}}
\,\,\,\,\,\,\,\,\,\,\,\,\,\,\,\,\,\,\,\,\,\,\,\,\,\,\,\,
\tilde B\,=\,(\vec\rho\cdot\vec\phi
\,+\,\vec\rho^{\,\,\dagger}\cdot\vec\phi^{\,\,\dagger}\,)\,
e^{-\vec \mu\cdot\vec\phi\,-\,
\vec\mu^{\,\,\dagger}\cdot\vec\phi^{\,\,\dagger}}.
\eqn\xseis
$$
In eq. \xseis\
$\vec\mu$, $\vec\mu^{\,\,\dagger}$, $\vec\rho$ and
$\vec\rho^{\,\,\dagger}$ are numerical vectors to be determined.
Requiring  the new fields to behave as in eqs. \vuno\ and \xcinco, one
gets the form of the remaining members of the multiplets, together
with many conditions that the numerical vectors must satisfy. These
conditions are enough to determine
$\vec\mu$, $\vec\mu^{\,\,\dagger}$, $\vec\rho$ and
$\vec\rho^{\,\,\dagger}$. Using the parametrization of $Q_M$ and $Q_L$
given in eq. \xcuatro, the final expressions of the fields
$(C, \Gamma^+, \Gamma^-, \tilde C)$ are:
$$
\eqalign{
C =& \,e^{{i\l}(\fm +\fmc-i\fl-i\flc)} \cr
\Gamma^+ =& \,-{\l}(\pmc -i\plc)\, e^{{i\l}(\fm +\fmc-i\fl-i\flc)} \cr
\Gamma^- =& \,{\l}(\psm -i\psl)\, e^{{i\l}(\fm +\fmc-i\fl-i\flc)} \cr
{\tilde C} =& \,[ \l^2(\pmc -i\plc)(\psm -i\psl)
                +{i\l\over 2} (\dfm-\dfmc -i\dfl +i\dflc)]
\, e^{{i\l}(\fm +\fmc-i\fl-i\flc)}, \cr}
\eqn\xsiete
$$
while the conjugate fields are given by:
$$
\eqalign{
{\tilde B} =& \, {i\over {2\l}} (\fm-\fmc+i\fl-i\flc)
            \, e^{-{i\l}(\fm +\fmc -i\fl -i\flc)} \cr
{\cal B}^+ =& \,[{1\over {2\l}} (\pmc + i\plc) +
{i\over 2} (\pmc-i\plc)(\fm-\fmc+i\fl-i\flc)]
              \, e^{-{i\l}(\fm +\fmc-i\fl-i\flc)} \cr
{\cal B}^- =& \,[ -{1\over {2\l}}(\psm + i\psl) +
{i\over 2} (\psm-i\psl)(\fm-\fmc+i\fl-i\flc)]
              \, e^{-{i\l}(\fm +\fmc -i\fl-i\flc)} \cr
B =& \,[ i\,{\l\over2} (\pmc-i\plc) (\psm -i\psl)(\fm-\fmc+i\fl-i\flc)
\cr & \;\;-i\pmc\psl -i\psm\plc + { i\over {4\l}}(\dfm +\dfmc +i\dfl
+i\dflc) \cr & \;\;+{1\over 4}(\dfm-\dfmc-i\dfl+i\dflc)
(\fm -\fmc +i\fl- i\flc)] \, e^{-{i\l}(\fm +\fmc-i\fl-i\flc)}. \cr}
\eqn\xocho
$$
It is straightforward, although in some cases tedious, to prove that the
matter+Liouville contributions to $T$, $T_F^{\pm}$ and $J$ can be
written as:
$$
\eqalign{
T^M\,+\,T^L\,=&\,2B\partial C\,+\,\partial B C\,-\,
{3\over 2}\, {\cal B}^+\partial\Gamma^-\,
-\,{1\over 2}\partial{\cal B}^+\,\Gamma^-\,-\,
{3\over 2}\,{\cal B}^-\partial\Gamma^+\,-\,
{1\over 2}\,\partial{\cal B}^-\Gamma^+\,+\,
\tilde B\partial\tilde C\cr
T_F^{ M,\pm}\,+\,T_F^{ L,\pm}\,=&\,
\mp B\Gamma^{\pm}\,+\,\partial\Gamma^{\pm}\tilde B\,+
{1\over 2}\, \Gamma^{\pm}\partial\tilde B\,\pm\,
{3\over 2}\, {\cal B}^{\pm}\partial C\,\pm\,
\partial{\cal B}^{\pm} C\,+\,\tilde C{\cal B}^{\pm}\cr
J^M\,+\, J^L\,=&\,{\cal B}^+\Gamma^-\,-\,{\cal
B}^-\Gamma^+\,+\, \partial(\tilde B C).\cr}
\eqn\xnueve
$$
As in the $N=1$ case, the topological symmetry relating the ghost and
matter+Liouville sectors is now evident. We can take
the current of its generator as:
$$
Q\,=\,bC\,+\,\tilde b \tilde C-\, \beta_+\Gamma^-\,-\,\beta_-\Gamma^+.
\eqn\xdiez
$$
Moreover, it is also possible to fulfill eq. \dseis\ with $d=0$ where $G$
and $R$ are given by:
$$
\eqalign{
G\,=&\,c\partial B\,+\,2\partial c B\,-\,
{1\over 2}\,\gamma_+\partial {\cal B}^-\,-\,
{3\over 2}\, \partial\gamma_+{\cal B}^-\,-
{1\over 2}\,\gamma_-\partial {\cal B}^+\,-\,
{3\over 2}\, \partial\gamma_-{\cal B}^+\cr
R\,=&\,cb\,+\,2BC\,+{1\over 2}\,\beta_+\gamma_-\,
+{1\over 2}\,\beta_-\gamma_+\,-{3\over 2}\,{\cal B}^+\Gamma^-\,
\,-{3\over 2}\,{\cal B}^-\Gamma^+\,+\tilde B\tilde C.\cr}
\eqn\xonce
$$
Interestingly, with respect to the $R$ current of eq. \xonce, the $N=2$
supersymmetry generators of eqs. \xtres\ and \xnueve\ split in a way
very similar to the $N=1$ case. Indeed one can easily check that the
terms in $T_F^{\pm}$ that contain derivatives of the fields have
$R$-charge $+1/2$ whereas those without derivatives have $R$-charge
$-1/2$. Therefore, writing $T_F^+\,=\,T_F^{++}\,+\,T_F^{+-}$ and
$T_F^-\,=\,T_F^{-+}\,+\,T_F^{--}$,  where $T_F^{\pm +}$ ($T_F^{\pm -}$)
have $R$-charge $+1/2$ ($-1/2$), one has:
$$
\eqalign{
T_F^{\pm +}\,=&\,\partial\Gamma^{\pm}\tilde B\,+
{1\over 2}\, \Gamma^{\pm}\partial\tilde B\,\pm\,
{3\over 2}\, {\cal B}^{\pm}\partial C\,\pm\,
\partial{\cal B}^{\pm} C\,+\,\partial\gamma_{\pm}\tilde b\,+\,
{1\over 2}\,\gamma_{\pm}\partial\tilde b\,\pm\,{3\over 2}\,
\beta_{\pm}\partial c\,\pm\partial\beta_{\pm}c\cr
T_F^{\pm -}\,=&\,\mp B\Gamma^{\pm}\,+\tilde C{\cal B}^{\pm}\,
\mp\, b\gamma_{\pm}\,+\tilde c\beta_{\pm}.\cr}
\eqn\xdoce
$$
Moreover, the $N=2$ current $J$ splits in a similar way. It can be
checked that $T_F^{\pm\pm}$ close a superconformal algebra with twelve
generators. We have thus obtained an extension of
the supersymmetry induced by the
topological symmetry for the $N=2$ string , which is
completely similar to the splitting found for the $N=1$ NSR string.
Actually, one can regard this supersymmetry as the hidden $N=4$
supersymmetry of the ghost sector of the $N=2$ superstring. We will
not attempt to study this supersymmetry here. Let us only mention that,
curiously, it can also be related to the small $N=4$ superconformal
algebra.

In conclusion, we have found a bosonization that allows to give a
topological supergravity representation of the
energy-momentum tensor and supersymmetry generators of the non-critical
superstrings with minimal matter. The compatibility of the topological
symmetry and supersymmetry induces a ``doubling" of the latter. Many
aspects of this bosonization remain to be explored. It would be
interesting, for example, to characterize the physical states of the
superstring that can be obtained as local excitations of our topological
supergravity fields. In view of the results of [\LlatasRoy] for the
bosonic string, we expect to recover in this supergravity representation
only a part of the physical state spectrum of the non-critical
superstrings. We intend to study this and other related issues in the
future.

\ack

We would like to thank I.P. Ennes, J. M. F. Labastida and P.M. Llatas
for helpful discussions.   This work was partially
supported  by DGICYT under grant PB 93-0344, by CICYT under
grant AEN 94-0928 and by a Spanish Ministry of Education (MEC)
fellowship for one of us (S. R.).

\refout

\end